\documentclass[journal=jacsat,manuscript=article]{achemso}

\usepackage[version=3]{mhchem} 

\usepackage{graphicx}

\usepackage{pdflscape}
\usepackage{ulem}

\usepackage{amsmath}
\usepackage{longtable}

\usepackage{xcolor}

\newcommand{\halpha}{H$\alpha$}

\newcommand{\htwo}{H$_2$}
\newcommand{\htwoo}{H$_2$O}
\newcommand{\otwo}{O$_2$}

\newcommand{\op}{O$^+$}

\newcommand{\cotwo}{CO$_2$}

\newcommand{\hei} {He {\sc i}}

\newcommand{\oones}{O($^1$$S$)}

\author{Antonio Garc\'ia Mu\~noz}
\email{antonio.garciamunoz@cea.fr}
\author{Ewan Bataille}
\affiliation[]{Universit\'e Paris-Saclay, Universit\'e Paris Cit\'e, CEA, CNRS, AIM, 91191, Gif-sur-Yvette, France}

\title[An \textsf{achemso} demo]
  {On the heating, excitation, dissociation and ionization of molecules  by high-energy photons in planetary atmospheres.
  }

\abbreviations{IR,NMR,UV}

\begin{document}

\begin{abstract}
Photoionization by 
high-energy photons creates non-thermal electrons with a broad range of energies that heat and chemically transform the atmospheres of planets. The specifics of the interactions are notably different when the gas is 
 atomic or molecular. Motivated by the idea that molecules survive to high altitude in some exoplanets, we built a model for the energy transfer from non-thermal electrons to the {\htwoo}, {\htwo} and {\otwo} molecules. Our calculations show that the primary electrons of energy above about a hundred eV, a likely outcome from X-ray photoionization at moderately high atmospheric densities, expend most of their energy in ionization, dissociation and electronic excitation collisions. 
\textcolor{black}{
In contrast, the primary electrons of less than about ten eV, such as those produced by extreme-ultraviolet photons at low densities, expend most of their energy in momentum transfer (heating), rotational and vibrational excitation collisions.
The partitioning between channels of weak thresholds is particularly sensitive to the local fractional ionization.} The transition between these two situations introduces a parallel transition in the way the stellar energy is deposited in the atmosphere.
Our calculations show that the 
 non-thermal electrons enhance the ionization rate by a factor of a few or more with respect to photoionization alone, but may not greatly contribute to the direct dissociation of molecules unless the local flux of far-ultraviolet photons is relatively weak. 
 These findings highlight the importance of tracking the energy from the incident photons to the non-thermal electrons and on to the gas for problems concerned with the remote sensing and energy balance of exoplanet atmospheres.

\end{abstract}

\section*{Keywords}
Exoplanet atmosphere, steam atmosphere, non-thermal electrons, energy transfer, energy degradation, Monte Carlo

\section{Introduction}

Photoionization is a fundamental process in planetary atmospheres. The primary electrons 
ejected by the incident high-energy photons transfer their (kinetic) energy to the background atmosphere through collisions, thereby creating secondary electrons as well as exciting and dissociating other atoms and molecules. The energy of the primary electrons is 
$E_0${$\sim$}$hc$/$\lambda${$-$}IP($X$),  $\lambda$ being the photon wavelength and IP($X$) the ionization potential of the atom or molecule $X$. 
For example, 
the H and O atoms photoionize at $\sim$911 {\AA} but the corresponding threshold for
{\htwoo} is only 984 {\AA}. 
Reference $E_0$ values produced by X-ray ($<$100 {\AA}), extreme-ultraviolet (EUV; 100-911 {\AA}) and far-ultraviolet (FUV; say 911-2000 {\AA}) \textcolor{black}{photons} are on the order of 
a few hundred eV, a few tens of eV and 
 a fraction of an eV, respectively. 
Because the photoionization cross sections typically decay towards short wavelengths, the primary electrons in the uppermost atmospheric layers where the stellar photons first interact with the atmosphere have energies of a few eV or less. They engage in inelastic collisions having weak thresholds, such as rotational, vibrational and possibly some electronic excitations. In contrast, the primary electrons formed in deep atmospheric layers by X-ray photoionization 
have energies of hundreds of eV or more and can dissociate and ionize further the gas.
\\

The transition between the two layers is not clear-cut and depends on the local composition and fractional ionization of the gas and on the stellar radiation. 
As molecules have collisional channels (rotational and vibrational excitation, dissociation) that are unavailable to atoms, the details of the transition are sensitive to whether the gas remains atomic or molecular. 
\textcolor{black}{Mapping the transition onto pressure in the atmosphere depends on factors such as the planet's gravity and the gas metallicity. Stellar photons of a given
(say X-ray) wavelength are deposited where the atmospheric opacity $\tau_{X}${$\sim$}1, which occurs at the pressure $p_{\tau_X \sim 1}${$\sim$}{$\mu$}{$g$}/$\sigma_{X}$, where $\mu$ is the mean molecular weight of the gas and $\sigma_{X}$ its photoionization cross section, and 
$g$ the planet's gravitational acceleration.\cite{garciamunozetal2024} 
The dependence on metallicity arises from both $\mu$ and $\sigma_{X}$ with the latter probably making more of a difference especially at short wavelengths.
For example,   
 $\sigma_{X}${$\sim$}10$^{-20}$ cm$^2$ at 100 {\AA} for an atmosphere dominated by {\htwo}-H, but {$\sim$}10-100 times larger if the gas is dominated by O, {\htwoo} or {\otwo}.
}
\\

There is a vast literature concerned with non-thermal electrons in the atmospheres of the Solar System planets\cite{fox2006,wedlundetal2011} 
and comets\cite{bhardwaj2003}. In them, the electrons typically interact with a neutral gas. 
This condition is important as
low fractional ionizations tend to minimize the  energy transferred from the primary electrons to the thermal electrons and therefore to heat.
\\

The study of non-thermal electrons in exoplanet atmospheres remains limited and mostly focused on hydrogen-dominated planets. 
\citet{cecchi-pestellinietal2006,cecchi-pestellinietal2009} and \citet{shematovichetal2014} estimated the fraction of the primary electrons' energy that goes into heat, showing that it can be substantially smaller than one especially at high energies and where the gas remains neutral. 
\citet{guoben-jaffel2016} adopted published prescriptions for the ionization and heating in {\htwo}/H/He mixtures to explore the 
energy transferred to the  gas under various irradiation conditions. 
\textcolor{black}{\citet{loccietal2022} investigated the ion-neutral chemistry triggered by the primary and secondary electrons produced in X-ray photoionization, showing that the abundance of some molecules is sensitive to their combined effect.}
In their models, the primary electrons are released mostly from  metals (atoms and molecules heavier than helium), but once released they slow down through collisions with the dominant 
{\htwo}/H/He. 
\citet{garciamunoz2023_icarus} discussed the role of non-thermal electrons for 
\textcolor{black}{both populating and depopulating}
the excited states of the H atom that are probed with transmission spectroscopy in the {\halpha} line, and proposed that similar ideas hold valid for the metastable He state that is probed in 
the {\hei} triplet line at 1083 nm.
\citet{gilletetal2023}
confirmed and extended the \citet{guoben-jaffel2016} findings for atmospheres of atomic hydrogen. 
\\

The expected diversity of atmospheric compositions and stellar irradiation conditions at exoplanets anticipates a complex role of non-thermal electrons. 
\citet{johnstoneetal2018}, \citet{nakayamaetal2022} and \citet{yoshidaetal2022}
implemented approximate methods to explore the role of non-thermal electrons in {\cotwo} and {\htwoo} atmospheres in their hydrodynamical calculations. It is difficult to assess from them though the fate of each collision product and their contribution
to the energy budget or to the chemistry. 
\citet{garciamunoz2023_aa} used an MC scheme to explore the energy transferred from the primary electrons to the H and O atoms of an atmosphere, but did not explore the transfer to the parent molecules. 
Clearly, there is a need to elucidate 
with methods that can be considered exact
how the energy transfer occurs in 
general atmospheres 
where the non-thermal electrons interact with multiple atoms and molecules.
\\

\section{Setting of this work}

\begin{figure*}
 \includegraphics[width=15cm]{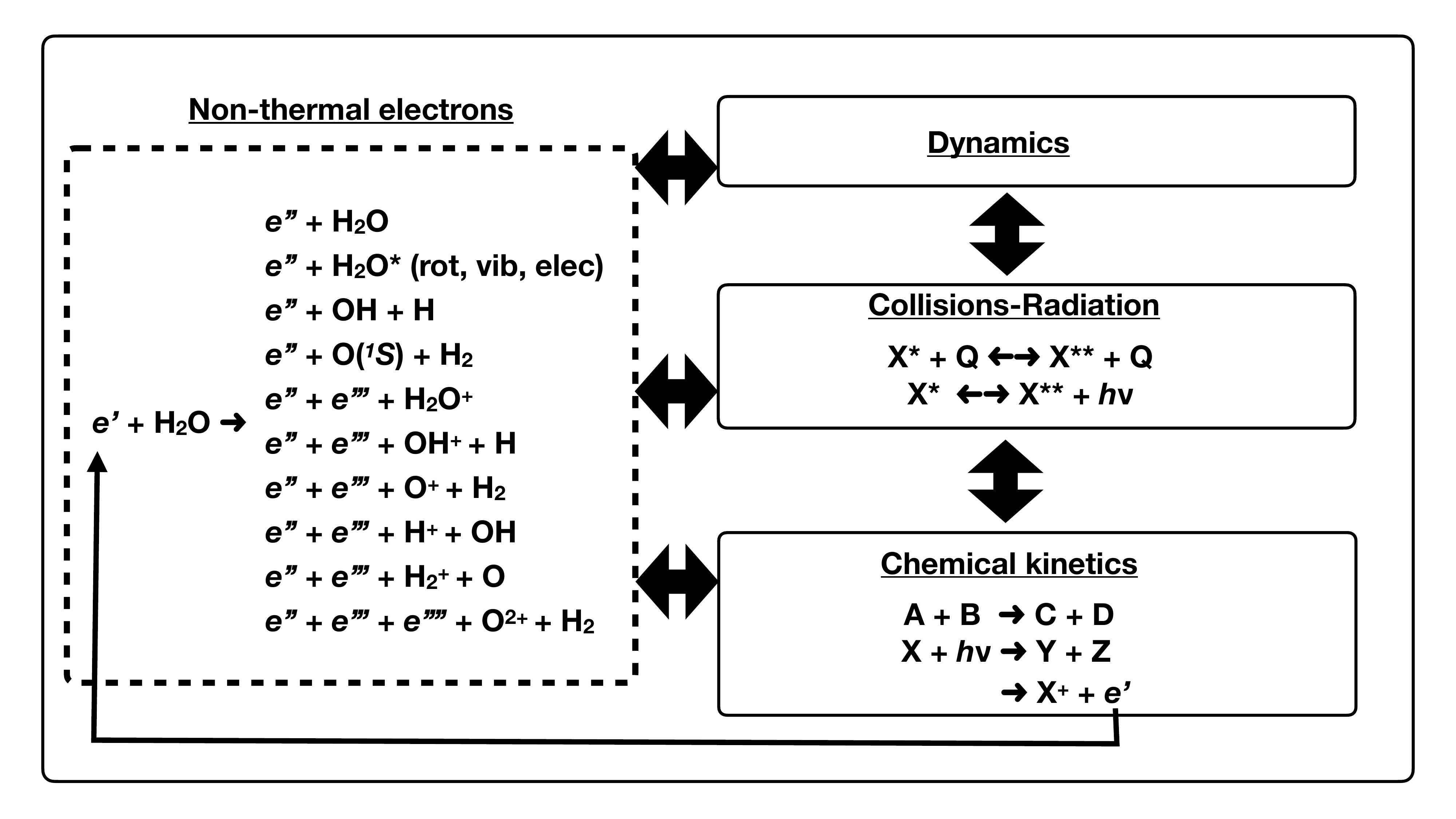}\\
 \caption{The current work treats the partial problem of non-thermal electrons
 (left box), with an emphasis on molecule-rich atmospheres ({\htwoo}-rich in the example). This 
  is part of a more general problem that connects the dynamics, collisional, radiative and 
 chemical kinetics of the atmospheric gas.
 }
 \label{partialproblem_fig}
\end{figure*}

Our work is concerned with the partial problem of understanding how the primary electrons created by  photoionization expend their  energy from creation to thermalization
 and, therefore, with determining whether the collisions they undergo result initially in the  heating, excitation, dissociation or ionization of the atmospheric gas. The problem is embedded in the more general problem of tracking the collision products through the full range of reactive, collisional and radiative interactions plus atmospheric dynamics. A full description of the general problem is essential for constraining the energy budget of the atmosphere and the abundances of the atoms and molecules targeted by  remote sensing techniques. 
\textcolor{black}{
Our current focus on the partial problem sets valuable guidance towards the general problem, 
which will be addressed in follow-up work, 
while extending past detailed work\cite{cecchi-pestellinietal2006,cecchi-pestellinietal2009,shematovichetal2014} to conditions in which the non-thermal electrons are created and slowed down by a variety of gases.
}
 Figure \ref{partialproblem_fig} sketches 
the partial and general problems.\\

\newpage

We focus on atmospheres that are rich in the {\htwoo}, {\htwo} and {\otwo} molecules. The motivation comes from the idea that such atmospheres are common \citep{venturinietal2020,kimuraikoma2022} \textcolor{black}{in exoplanets} and may represent the conditions of early Earth, Venus and Mars. Under some conditions\cite{garciamunoz2023_aa}, the molecules survive to high altitudes, where they become ionized and dissociated by high-energy stellar photons. Foreseably, the same molecules act as both donors of primary electrons to the background gas and as the dominant collisional targets to slow them down. 
\\

The work is split in two blocks. The first one describes the Monte Carlo (MC) scheme used to solve the slowing down problem of the non-thermal electrons. We list some of the production yields calculated for each of the {\htwoo}, {\htwo} and {\otwo} molecules and compare them to the information available in the literature. In the second block, we apply our method to a set of published atmospheric profiles\citep{garciamunoz2023_aa} for the exoplanet Trappist-1 b. 
\textcolor{black}{These profiles contain, amongst other gases, thermal electrons, the H, O, {\op} atoms and the
{\htwoo}, {\htwo} and {\otwo} molecules. These particles participate in the slowing down of the non-thermal electrons.}
We discuss the characteristic energy of the primary electrons that are released and how much of their energy goes into heating, excitation, dissociation and ionization following the elastic and inelastic collisions implemented in the MC scheme. Lastly, we discuss the contribution of secondary electrons to further ionization and dissociation.

\section{The slowing down problem}

The non-thermal electrons lose their kinetic energy in collisions with the thermal electrons, atoms and molecules in the gas.
We solve the
slowing down problem for the non-thermal electrons with a MC scheme originally devised for collisions with atoms. \citep{garciamunoz2023_icarus,garciamunoz2023_aa} The scheme is extended here to accommodate  
{\htwoo}, {\htwo} and {\otwo} and some forms of inelastic collisions (dissociative attachment, rotational and vibrational excitation, dissociation) not available to atoms.

\subsubsection{Collisions with {\htwoo}}
\label{h2o_implementation_MC_section}

Table \ref{h2ochannels_table} lists the channels for collisions of the non-thermal electrons with {\htwoo} implemented in the MC scheme. The channels and 
energies of the {\htwoo} states
are from \citet{songetal2021} 
The description of
electronic excitation (EE) remains problematic,
\citep[][]{xuetal2021} but it is believed to result in dissociation with high efficiency. We merged the EE channels 
$\rightarrow$${\rm{H}}_2{\rm{O}}(\tilde{a}^3B_1,\tilde{A}^1B_1,^3A_2,^1A_2,\tilde{b}^3A_1,\tilde{B}^1A_1)$, 
on which there is partial experimental and theoretical information, 
into a single channel $\rightarrow$${\rm{H}}_2{\rm{O}^*}$ 
that connects the ground electronic state with a super-state of energy 8.4 eV.
We adopted for this EE channel the total of the cross sections measured by \citet{ralphsetal2013} up to 20 eV. 
\textcolor{black}{
For channel DN1 concerned with dissociation into neutrals, we adopted the cross sections from 
\citet{itikawamason2005h2o} and, to avoid double counting, subtracted from them the EE cross sections.
}
We favored for DN1 the 
dissociation cross sections in \citet{itikawamason2005h2o} over those in \citet{songetal2021} 
because the former result in a somewhat better match of the MC calculations to the experimental ionization yields
(see below). 
We extrapolated the cross sections for ionization
(channels IO1-IO6) and dissociation into OH$+$H (DN1) to high energies 
with a law of the type $\sigma${$\propto$}$\ln{E}$/$E$.
\textcolor{black}{This law is theoretically supported for ionization and excitation in dipole-allowed transitions in atoms \citep{ralchenkoetal2008}. It seems to reproduce well the behavior for ionization and dissociation in molecules when such data are available.} 
\textcolor{black}{Dissociative attachment, channels DA1-DA3, create anions with the potential of triggering further negative ion chemistry.\citep{shumanetal2015} Our calculations show however that these channels are relatively minor with respect to other channels that have also weak thresholds and we omit their discussion.}
\\

\textcolor{black}{
Our calculations assume the
target {\htwoo} molecule in its vibrational and electronic ground state, a reasonable assumption at temperatures less than say 2000 K. 
It would be interesting to explore how the vibrational excitation of the molecule might change some of the quantitative conclusions at higher temperatures, but that requires sets of inelastic cross sections for the vibrationally excited states that are mostly unavailable.
}
The treatment in the MC scheme of momentum transfer (MT), electronic excitation 
and ionization is analogous to the treatment for atoms.\citep{garciamunoz2023_icarus}  
For the partitioning of energy between the fast and slow electrons in ionization collisions, 
we used analytical probabilities with 
$\bar{E}$=13 eV.\citep{opaletal1971,garciamunoz2023_icarus} 
In the treatment of all channels except 
dissociative attachment (DA1-DA3), 
the MC scheme
assumes that the scattered and ejected electrons carry away the excess energy or, equivalently, that the post-collision heavy particles have zero kinetic energy. 
Unfortunately, the information available on the energy transferred to the heavy particles is scarce
and does not allow a more realistic treatment. For completeness, we note that the available dissociation data suggest that this simplification may introduce systematic errors \citep{makarovetal2004}, and
may increase the calculated yields over their true values because it extends the number of collisions undergone by the non-thermal electrons.  
Channels DA1-DA3 are treated by 
forcing the incident electron to a full stop, which entails that the excess kinetic energy is transferred to the fragments. 
\\

The slowing-down of the electrons in purely rotational collisions (channel RE)
occurs simultaneously over many rotational states. 
The rotational stopping cross section 
$
S_{\rm{rot}}(E) = \sum_i f_i \sum_j \sigma_{{\rm{rot}},i \rightarrow j}(E) {\times}
(E_j - E_i)
$ [cm$^2$eV]
is defined by the summation over all rotational states $i$ of their relative abundance $f_i$ ($\sum_i f_i${$\equiv$}1)
times the rotational cross section for the transition from state $i$ to $j$ (excitation if $E_j${$>$}$E_i$; deexcitation 
if $E_j${$<$}$E_i$) times the energy difference between the states.
Being a stopping cross section, 
it conveys the net energy 
transferred by the incident electrons per unit of {\htwoo} column traversed.
The summation over $j$ is insensitive (or weakly sensitive) to the 
actual state $i$.\citep{shimamura1984,shimamura1990,itikawamason2005rot} 
Therefore $S_{\rm{rot}}$ is insensitive
to the population details for the rotational states. 
We calculated $S_{\rm{rot}}$ 
with the cross sections for $J$=0{$\rightarrow$}$J$=1 in
\citet{songetal2021} and 
$E_{J=1}${$-$}$E_{J=0}$=4.6$\times$10$^{-3}$ eV. 
The MC scheme requires the prescription of both a 
characteristic cross section {$<$}$\sigma_{\rm{rot}}${$>$}
and energy {$<$}$\Delta E_{\rm{rot}}${$>$} lost by the electron at each collision.  We adopted 
{$<$}$\Delta E_{\rm{rot}}${$>$}=4.6$\times$10$^{-3}$ eV and
{$<$}$\sigma_{\rm{rot}}${$>$}={$S_{\rm{rot}}$}/{$<$}$\Delta E_{\rm{rot}}${$>$}. 
This pseudo-continuous approach is akin to the way in which 
the slowing-down of non-thermal electrons in collisions with thermal electrons, itself a continuous process, is treated.\citep{garciamunoz2023_icarus} 
\\

\begin{table}
\begin{small}
 \caption{\small{Channels for the slowing down of non-thermal electrons in collisions with {\htwoo}.
 For rotational excitation (RE), $J_{\tau}$ refers to the quantum numbers that describe the 
 rotational motion. 
 For vibrational excitation (VE), the numbers in parenthesis refer to the quantum numbers $v$=($v_1v_2v_3$) for symmetric stretching ($v_1$), bending ($v_2$) and asymmetric stretching ($v_3$). 
In the MC scheme, 
 both stretching modes are lumped into one channel.
 For electronic excitation (EE), the symbol in parenthesis refers to the ground electronic state, and {\htwoo}$^*$ to the proposed super-state. 
 Some information about the channels for momentum transfer (MT), dissociation into neutrals (DN), ionization (IO) and dissociative attachment (DA) is given in the text.
 The column before the last quotes the energy extracted 
 from the electrons at each collision in the MC scheme, with $E'$ standing for the incident electron energy.
The last column describes how each channel may contribute to the macroscopic chemical kinetics of the general problem (see Fig. \ref{partialproblem_fig}).}
 }
 \label{h2ochannels_table}
\rotatebox{90}{
 \begin{tabular}{lccccc}

\hline
  Channel & Pre-collision &  & Post-collision & Energy extracted  & Chemical modelling: {\htwoo}{$\rightarrow$} \\  
\hline
  MT & $e' + {\rm{H}}_2{\rm{O}}$ & $\rightarrow$ & $e'' + {\rm{H}}_2{\rm{O}}$ & (2$m_e$/$m_{{\rm{H}}_2{\rm{O}}}$){$E'$} & {\htwoo} \\
  \hline
  RE & $e' + {\rm{H}}_2{\rm{O}}(J'_{\tau'})$ & $\rightarrow$ & $e'' + {\rm{H}}_2{\rm{O}}(J''_{\tau''})$ & 0.0046 eV & {\htwoo} \\ 
  \hline
  VE2  & $e' + {\rm{H}}_2{\rm{O}}(000)$ & $\rightarrow$ & $e'' + {\rm{H}}_2{\rm{O}}(010)$ & 0.198 eV & {\htwoo} \\      
  VE13 &  & $\rightarrow$ & $e'' + {\rm{H}}_2{\rm{O}}(100,001)$ & 0.46 eV & {\htwoo} \\  
  \hline  
EE & $e' + {\rm{H}}_2{\rm{O}}(X ^1A_1)$ & $\rightarrow$ & $e'' + {\rm{H}}_2{\rm{O}}^*$ & 8.4 eV & {OH}$+$H\\    
\hline
DN1 & $e' + {\rm{H}}_2{\rm{O}}$ & $\rightarrow$ & $e'' + {\rm{OH}} + {\rm{H}}$ & 5.11 eV & OH{$+$}H \\ 
DN2 &                           & $\rightarrow$ & $e'' + {\rm{O}}(^1S) + {\rm{H}_2}$ & 9.21 eV & {\oones}$+${\htwo} \\  
  \hline  
IO1 & $e' + {\rm{H}}_2{\rm{O}}$ & $\rightarrow$ & $e'' + e''' + {\rm{H}}_2{\rm{O}}^+$ & 12.61 eV & ${\rm{H}}_2{\rm{O}}^+${$+$}$e^-$\\    
IO2 &                           & $\rightarrow$ & $e'' + e''' + {\rm{OH}^+} + {\rm{H}}$ & 18.11 eV & ${\rm{OH}}^+${$+$}H{$+$}$e^-$\\ 
IO3 &                           & $\rightarrow$ & $e'' + e''' + {\rm{O}^+} + {\rm{H}_2}$ & 18.64 eV & ${\rm{O}}^+${$+$}H$_2${$+$}$e^-$\\    
IO4 &                           & $\rightarrow$ & $e'' + e''' + {\rm{H}}^{+} + {\rm{OH}}$ & 18.72 eV & ${\rm{H}}^+${$+$}OH{$+$}$e^-$\\    
IO5 &                           & $\rightarrow$ & $e'' + e''' + {\rm{H}_2^+} + {\rm{O}}$ & 20.46 eV & {H$_2^+$}{$+$O}{$+$}$e^-$\\    
IO6 &                           & $\rightarrow$ & $e'' + e''' + e'''' + {\rm{O}^{2+}} + \rm{H}_2$  & 53.76 eV & O$^{2+}${$+$H$_2$}{$+$}$e^-${$+$}$e^-$ \\    
  \hline  
DA1 & $e' + {\rm{H}}_2{\rm{O}}$ & $\rightarrow$ & ${\rm{OH}^-} + {\rm{H}}$ & $E'$ & {\htwoo} \\  
DA2 & & $\rightarrow$ & ${\rm{O}^-} + {\rm{H}_2}$ & $E'$ & {\htwoo} \\  
DA3 &  & $\rightarrow$ & ${\rm{H}^-} + {\rm{OH}}$ & $E'$ & {\htwoo} \\  

\hline
\end{tabular}
}
\end{small}
\end{table}

\subsubsection{Collisions with {\htwo}}
\label{h2_implementation_mc_section}

Table \ref{h2channels_table} lists the channels for collisions with {\htwo}. 
We relied for the compilation of channels and cross sections mainly on \citet{yoonetal2008}
For the energies, we used 
\citet{fantzwunderlich2006}
for the bound states, \citet{sharp1971} for the vertical energies connecting with the repulsive $b$ state, and the  KIDA\citep{wakelametal2012} and UMIST\citep{mcelroyetal2013} databases for ionization. 
\textcolor{black}{
 We adopted $\bar{E}$=8.3 eV for the partitioning between the fast and slow electrons following ionization.\citep{opaletal1971}}
We formed $S_{\rm{rot}}$ with the excitation cross sections for $J$=0{$\rightarrow$}$J$=2 and the corresponding energy between states, and adopted the pseudo-continuous approach with {$<$}$\Delta E_{\rm{rot}}${$>$}=44$\times$10$^{-3}$ eV 
($\approx${$E_{J=2}$}$-${$E_{J=0}$}).
We consider vibrational excitation
($v$=0){$\rightarrow$}($v$=1) but omit the much weaker excitation into
($v${$>$}1). We omit vibrational excitation from ($v${$\ge$}1)
because under typical conditions 
the abundances of these excited states are too low to make a difference.
We explored the significance of super-elastic collisions of 
the non-thermal electrons leading to
($v$=1){$\rightarrow$}($v$=0), which occur with no energy threshold.  
Our calculations showed that they are not dominant for 
quenching (collisional deexcitation) of ($v$=1)
and that contribute negligibly to the return of energy to the non-thermal electrons, and we omit it too. 
We lumped the high-energy electronic states of {\htwo} into two super-states, $SS$ and $ss$ for the singlet and triplet manifolds, respectively, to which we assigned energies of 14 eV. 
We collected the cross sections 
of these high-energy states 
(summed over all final vibrational states) from \citet{scarlettetal2021} and added them for implementation in channels EE1SS and EE3ss.
The cross sections for
electronic excitation into singlet states and for ionization were extrapolated to high energies with a law of the type $\sigma${$\propto$}$\ln{E}$/$E$, and those for excitation into triplet states with 
one of the type
$\sigma${$\propto$}1/$E^3$. 
Dissociation into neutrals that results in the fragment H($2p$) is useful for diagnostic purposes. We consider it, with cross sections from \citet{ajelloetal1991}

\begin{table}
\begin{small}
 \caption{Similar to Table \ref{h2ochannels_table}, for collisions of non-thermal electrons with {\htwo}.
 }
 \label{h2channels_table}
 \rotatebox{90}{
 \begin{tabular}{lccccc}

\hline
  Channel & Pre-collision &  & Post-collision & Energy extracted  & Chemical modelling: {\htwo}{$\rightarrow$} \\  
\hline
  MT & $e' + {\rm{H}}_2$ & $\rightarrow$ & $e'' + {\rm{H}}_2$ & (2$m_e$/$m_{{\rm{H}}_2}$){$E'$} & {\htwo} \\
  \hline
  RE & $e' + {\rm{H}}_2(J')$ & $\rightarrow$ & $e'' + {\rm{H}}_2(J'')$ & 0.044 eV & {\htwo} \\ 
  \hline
  VE  & $e' + {\rm{H}}_2(v'=0)$ & $\rightarrow$ & $e'' + {\rm{H}}_2(v''=1)$ & 0.52 eV & {\htwo} \\      

\hline  

EE1B & $e' + {\rm{H}}_2(X^1\Sigma^+_g)$ & $\rightarrow$ & $e'' + {\rm{H}}_2(B^1\Sigma^+_u)$ & 11.18 eV & {\htwo}; H$+$H \\   

EE1C & $e' + {\rm{H}}_2(X^1\Sigma^+_g)$ & $\rightarrow$ & $e'' + {\rm{H}}_2(C^1\Pi_u)$ & 12.29 eV & {\htwo}; H$+$H \\   

EE1EF & $e' + {\rm{H}}_2(X^1\Sigma^+_g)$ & $\rightarrow$ & $e'' + {\rm{H}}_2(E,F^1\Sigma^+_g)$ & 12.30 eV & {\htwo}; H$+$H \\   

EE1SS & $e' + {\rm{H}}_2(X^1\Sigma^+_g)$ & $\rightarrow$ & $e'' + {\rm{H}_2}$($B'^1\Sigma_u^+, GK^1\Sigma_g^+,I^1\Pi_g,J^1\Delta_g, D^1\Pi_u, H\bar{H}^1\Sigma_g^+$) & 14 eV & {\htwo}; H$+$H \\   

  \hline  
 EE3b & $e' + {\rm{H}}_2(X^1\Sigma^+_g)$ & $\rightarrow$ & $e'' + {\rm{H}}_2(b^3\Sigma^+_u)$ & 10.45 eV & H$+$H \\   

 EE3c & $e' + {\rm{H}}_2(X^1\Sigma^+_g)$ & $\rightarrow$ & $e'' + {\rm{H}}_2(c^3\Pi_u)$ & 11.77 eV & H$+$H \\   

 EE3a & $e' + {\rm{H}}_2(X^1\Sigma^+_g)$ & $\rightarrow$ & $e'' + {\rm{H}}_2(a^3\Sigma^+_g)$ & 11.79 eV & 
 H$+$H \\   

 EE3e & $e' + {\rm{H}}_2(X^1\Sigma^+_g)$ & $\rightarrow$ & $e'' + {\rm{H}}_2(e^3\Sigma^+_u)$ & 13.22 eV & H$+$H \\   

EE3ss & $e' + {\rm{H}}_2(X^1\Sigma^+_g)$ & $\rightarrow$ & $e'' + {\rm{H}_2}$($d^3\Pi_u,h^3\Sigma_g^+,g^3\Sigma_g^+,i^3\Pi_g,j^3\Delta_g$) & 14 eV & H$+$H \\

\hline

 DN & $e' + {\rm{H}}_2$ & $\rightarrow$ & $e'' + {\rm{H}}(2p) + {\rm{H}}$ & 14.68 eV &
 ${\rm{H}}(2p) + {\rm{H}}$\\  

\hline
  
 IO1 & $e' + {\rm{H}}_2$ & $\rightarrow$ & $e'' + e''' + {\rm{H}}_2^+$ & 15.43 eV & ${\rm{H}}_2^+${$+$}$e^-$\\    
 IO2 &                           & $\rightarrow$ & $e'' + e''' + {\rm{H}}^{+} + {\rm{H}}$ & 18.08 eV & ${\rm{H}}^+${$+$}H{$+$}$e^-$\\    

  \hline  
 DA & $e' + {\rm{H}}_2$ & $\rightarrow$ & ${\rm{H}^-} + {\rm{H}}$ & $E'$ & {\htwo} \\  
\hline
\end{tabular}
}
\end{small}
\end{table}

\subsubsection{Collisions with {\otwo}}
\label{o2_implementation_mc_section}

Table \ref{o2channels_table} lists the channels for collisions with {\otwo}. 
We borrowed the channels and cross sections from \citet{itikawa2009}, the energies from this reference or from \citet{itikawaetal1989}, and
$\bar{E}$=17.4 eV from \citet{opaletal1971}
 We formed $S_{\rm{rot}}$ with the Born cross sections\citep{itikawa2009} for $J$=1{$\rightarrow$}$J$=3 
and the corresponding energy between states,
extracting 
{$<$}$\Delta E_{\rm{rot}}${$>$}=1.8$\times$10$^{-3}$ eV 
($\approx${$E_{J=3}$}$-${$E_{J=1}$})
from the incident electron at each collision.
DN includes all possible channels for dissociation into neutrals. 
We adopted the corresponding DN cross sections from \citet{suzukietal2011} 
(their Table IV), which 
compare favourably at small energies with the total 
cross sections for 
excitation into the Schumann-Runge continuum, the Longest Band and the Second Band in \citet{itikawa2009}, and also 
compare favourably with the total dissociation cross sections at high energies quoted in that reference. 
For energies at 
which no DN cross section measurements are available, 
we used the BEf representation from \citet{suzukietal2011}
and extrapolated it towards high energies with a $\sigma${$\propto$}$\ln{E}$/$E$ law.
At high energies, additional dissociation channels open up \citep{itikawa2009} that are omitted by \citet{suzukietal2011}  and therefore from our model. 
This omission is likely to underestimate our predictions for {\otwo} dissociation, in turn overestimating the 
predicted ionization yields. 
This deficiency may help explain the discrepancies between the calculated and measured ionization yields 
\textcolor{black}{(see discussion section below)}. 
We extrapolated the ionization cross sections towards high energies with a 
$\sigma${$\propto$}$\ln{E}$/$E$ law. 
\\

\begin{table}
\begin{small}
 \caption{\small{Similar to Table \ref{h2ochannels_table}, for collisions of non-thermal electrons with {\otwo}.}
 }
 \label{o2channels_table}
 \rotatebox{90}{
 \begin{tabular}{lccccc}
\hline
  Channel & Pre-collision &  & Post-collision & Energy extracted  & Chemical modelling: {\otwo}{$\rightarrow$} \\  
\hline
  MT & $e' + {\rm{O}}_2$ & $\rightarrow$ & $e'' + {\rm{O}}_2$ & (2$m_e$/$m_{{\rm{O}}_2}$){$E'$} & {\otwo} \\
  \hline
  RE & $e' + {\rm{O}}_2(J')$ & $\rightarrow$ & $e'' + {\rm{O}}_2(J'')$ & 0.0018 eV & {\otwo} \\ 
  \hline
  VE1  & $e' + {\rm{O}}_2(v'=0)$ & $\rightarrow$ & $e'' + {\rm{O}}_2(v''=1)$ & 0.193 eV & {\otwo} \\  
  
  VE2  & $e' + {\rm{O}}_2(v'=0)$ & $\rightarrow$ & $e'' + {\rm{O}}_2(v''=2)$ & 0.386 eV & {\otwo} \\ 
  
  VE3  & $e' + {\rm{O}}_2(v'=0)$ & $\rightarrow$ & $e'' + {\rm{O}}_2(v''=3)$ & 0.579 eV & {\otwo} \\

\hline  
EEa & $e' + {\rm{O}}_2(X^3\Sigma^-_g)$ & $\rightarrow$ & $e'' + {\rm{O}}_2(a^1\Delta_g)$ & 0.98 eV & {\otwo}($a$) \\   

EEb & $e' + {\rm{O}}_2(X^3\Sigma^-_g)$ & $\rightarrow$ & $e'' + {\rm{O}}_2(b^1\Sigma^+_g)$ & 1.63 eV & {\otwo} \\   

EEcA'A & $e' + {\rm{O}}_2(X^3\Sigma^-_g)$ & $\rightarrow$ & $e'' + {\rm{O}}_2(c^1\Sigma^-_u, A'^3\Delta_u, A^3\Sigma^+_u)$ & 4.05 eV & 
 {\otwo} \\   

\hline
DN & $e' + {\rm{O}}_2$ & $\rightarrow$ & $e'' + {\rm{O}}(^3P) + {\rm{O}}(^1D)$ & 7.07 eV & ${\rm{O}}(^3P) + {\rm{O}}(^1D)$\\  

\hline
IO1 & $e' + {\rm{O}}_2$ & $\rightarrow$ & $e'' + e''' + {\rm{O}}_2^+$ & 12.07 eV & $e' + {\rm{O}}_2^+$\\    

IO2 & $e' + {\rm{O}}_2 $ & $\rightarrow$ & $e'' + e''' + {\rm{O}}^{+}(^4S^{\rm{o}}) + {\rm{O}}$($^3P$) & 22.8 eV & $e' + {\rm{O}}^{+}(^4S^{\rm{o}}) + {\rm{O}}$($^3P$)\\    

IO3 & $e' + {\rm{O}}_2 $ & $\rightarrow$ & $e'' + e''' + e'''' + {\rm{O}}^{2+}(^3P) + {\rm{O}}$($^3P$) & 72 eV & $e' + e'' + {\rm{O}}^{2+}(^3P) + {\rm{O}}(^3P)$\\    

  \hline  
 DA & $e' + {\rm{O}}_2$ & $\rightarrow$ & ${\rm{O}^-} + {\rm{O}}$ & $E'$ & {\otwo} \\  
\hline
\end{tabular}
}
\end{small}
\end{table}

\subsubsection{Discussion. Comparison between molecules}

\begin{figure*}
 \includegraphics[width=15cm]{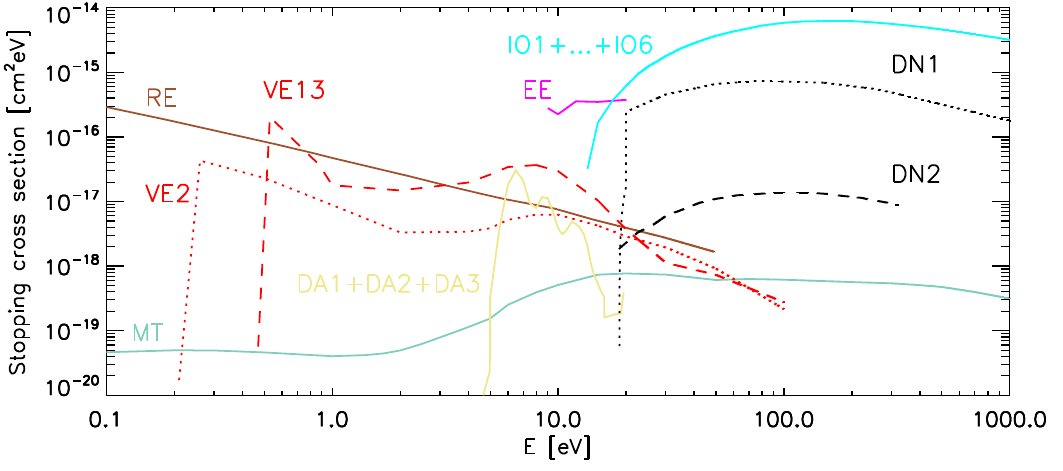}\\
 \includegraphics[width=15cm]{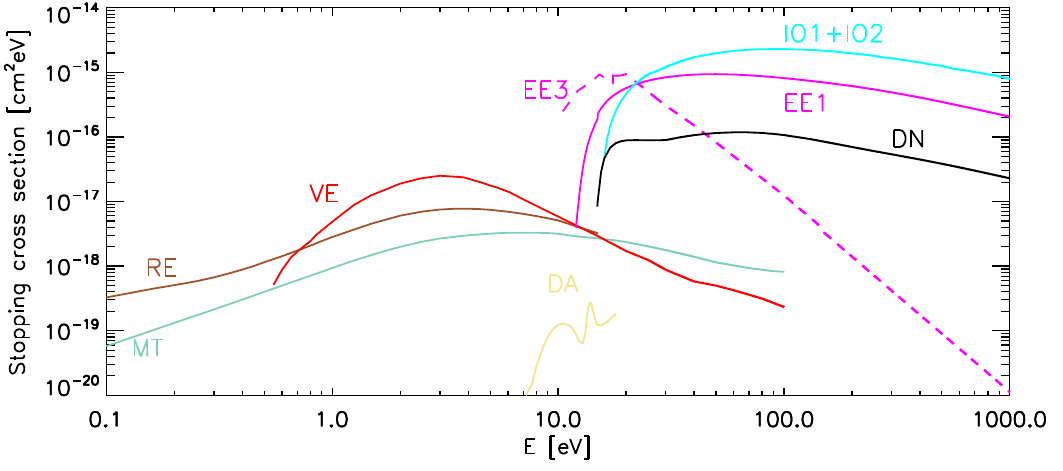}\\
 \includegraphics[width=15cm]{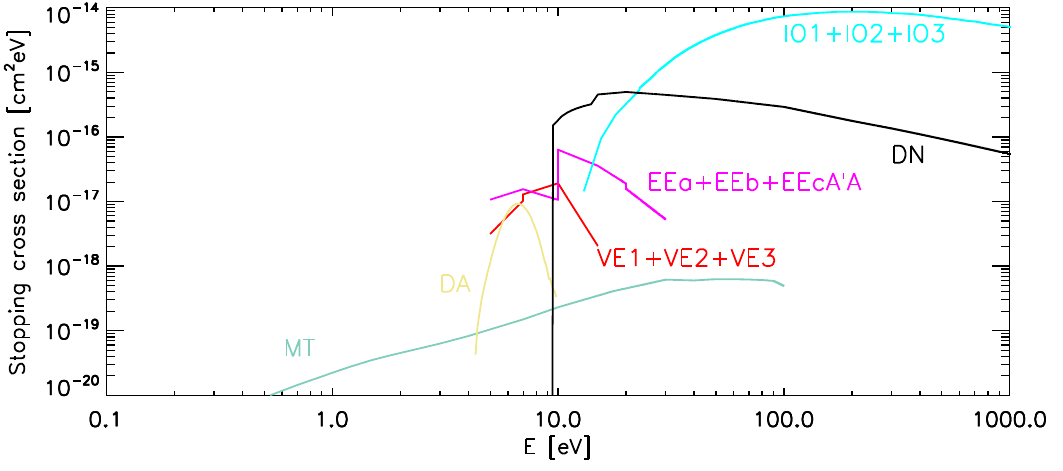}
 \caption{Stopping cross sections for collisions of non-thermal electrons with 
 {\htwoo} (top), {\htwo} (middle) and {\otwo} (bottom). Further information on the channels can be found in the text 
  and in Tables \ref{h2ochannels_table}-\ref{o2yields_table}.
In the middle panel, EE1 and EE3 represent the total of the stopping cross sections for electronic excitation into singlet and triplet states, respectively.
In the bottom panel, the RE channel falls below the 
range of values represented.
 }
 \label{stopxs_fig}
\end{figure*}

We formed stopping cross sections for each collisional channel, as they
offer a first glimpse into how the energy of the primary electrons is expended since they are released and until they are thermalized.
Figure \ref{stopxs_fig} 
shows the stopping cross sections for the channels in Tables \ref{h2ochannels_table}-\ref{o2channels_table}, 
calculated according to standard expressions
\citep{itikawa2007}.\footnote{For an ionization channel and an incident electron of energy $E'$, the stopping cross section is $S_{\rm{IO}}$($E'$)=$\sigma_{\rm{IO}}$($E'$)(IP+{$<$}$E''${$>$}), where IP is the ionization potential in the 
channel and {$<$}$E''${$>$} is the mean energy of the scattered electron (the slowest of the two post-collision electrons when  one electron is ejected). 
For probability distributions of the form $P$($E''; E'$){$\propto$1/(1+($E''$/$\bar{E}$)$^2$)} 
\citep{garciamunoz2023_icarus}, the mean energy takes on the analytical expression
{$<$}$E''${$>$}/$\bar{E}$=($\ln{\sqrt{1+x^2}}$)/$\arctan{x}$, where $x$=($E-\rm{IP}$)/(2{$\bar{E}$}).
For dissociative attachment, 
$S_{\rm{DA}}$($E'$)=$\sigma_{\rm{DA}}$($E'$)$E'$, as the incident electron is fully stopped.
} 
The {\htwoo} stopping cross sections are consistent  
with those reported at energies below 10 eV
by \citet{itikawa2007} (their Fig. 5.47). The {\htwo}
stopping cross sections 
are consistent with those in refs.
\cite{cravensetal1975,takayanagi1984,dalgarnoetal1999}.
The {\otwo} stopping cross sections match reasonably the calculations above 30 eV in ref.\cite{guptaetal1975}.
Figure \ref{stopxs_fig}
 suggests that 
ionization, dissociation and electronic excitation dominate the energy loss for high-energy primary electrons, and that 
for primary electrons of low-to-moderate energy, rotational and
vibrational excitation channels dominate the energy loss. The specifics are very molecule-dependent. 
\\

We  ran MC simulations in which primary electrons of energy $E_0$ are injected into a gas composed of a molecule
$X$(={\htwoo}, {\htwo}, {\otwo})
 and thermal electrons  
for fractional ionizations $x_e$=[$e_{\rm{th}}$]/[$X$] between 0 and 10$^{-2}$ (brackets denote number densities).
 Outcomes of such simulations are the production yields for each channel and the fraction of $E_0$ that goes into them.
Tables \ref{h2oyields_table}-\ref{o2yields_table} summarize some of the MC outputs. 
\textcolor{black}{
The complete set of production yields is available through the Supplementary Information (SI).}
For {\htwoo}, the $x_e$=0 yields are generally consistent with the vibrational yields of \citet{dayashankar1997} and the ionization yields of \citet{turneretal1982} and \citet{dayashankargreen1992}. 
We note that
our ionization yields at high energy are {$\sim$}15{\%} above the earlier calculations and the existent measurement
($\sim$1/30 eV$^{-1}$ for the number of ions/$E_0$; ref. \cite{combecher1980}).
 We believe 
the discrepancy arises from the implemented cross sections and, possibly, from assuming that no kinetic energy is transferred  in the collisions to the heavy particles.
To better understand this issue, 
 we performed some numerical experiments and confirmed that, for example, the excitation-vs-ionization partitioning is sensitive to the adopted DN1 cross sections at high energy, which suffer from notable uncertainties \citep{songetal2021}. 
 For {\htwo},
the comparison of the production yields calculated here 
with the calculations by \citet{cravensetal1975} (their figs. 2-5, 12-13) and \citet{dalgarnoetal1999} (figs. 6, 11-13)
is generally very good 
over a broad range of fractional ionizations.
For {\otwo}, 
our ionization yields are generally consistent with, yet somewhat lower than, past calculations\citep{greenetal1977}
in neutral {\otwo} 
and the experimental value of 
1/32.55=0.03072 eV$^{-1}$ at $E_0$=500 eV
\citep{combecher1980}. 
We attribute the discrepancies to similar reasons as for the {\htwoo} calculations 
(see also ref.\cite{makarovetal2003}). 
The yields for total vibrational excitation 
and excitation into $a^1\Delta_g$  and $b^1\Sigma^+_g$ 
for $E_0$=6 eV were calculated by \citet{ishiietal1992} 
They obtained 
4.10, 3.47 and 0.54 excitations/injected  electron, respectively. 
The yields calculated here are 1.97, 0.79, and 0.22. 
Considering the large uncertainties that exist in the relevant cross sections, 
we deem both sets of calculations acceptably consistent.
\\

\begin{table}
 \caption{Normalized yields in {\htwoo} or number of excitation, dissociation and ionization events divided by the non-thermal electron injection energy $E_0$ for a range of fractional ionizations of the gas. 
 For information on the channels, see Table \ref{h2ochannels_table}. 
 DN refers to the total of DN1+DN2, and  
 IO refers to the total of IO1+...+IO6. For comparison, \citet{combecher1980}
has measured normalized yields for ionization in neutral water of 1/35.15=0.02845 eV$^{-1}$ for $E_0$=100 eV and 1/30.38=0.03292 eV$^{-1}$
for $E_0$=900 eV \textcolor{black}{(to be compared directly with the yield at $E_0$=1,000 eV quoted here)}.
 }
 \label{h2oyields_table}
 \begin{tabular}{llccccc}

\hline
               &            & \multicolumn{5}{c}{Normalized yield [eV$^{-1}$]} \\
               \cline{3-7}
  $x_{\rm{e}}$ & $E_0$ [eV] & VE2 & VE13  & EE & DN & IO \\  
\hline  
   &    1. &   0.4980 &   0.5680 &   0.0000 &   0.0000 &   0.0000 \\
   &   10. &   0.2730 &   0.5850 &   0.0505 &   0.0000 &   0.0000 \\
0. &  100. &   0.0961 &   0.1850 &   0.0139 &   0.0364 &   0.0346 \\
   & 1000. &   0.0824 &   0.1600 &   0.0116 &   0.0353 &   0.0374 \\
   &10000. &   0.0816 &   0.1580 &   0.0115 &   0.0347 &   0.0380 \\  
\hline

         &    1. &   0.1560  &  0.1760 &   0.0000  &  0.0000  &  0.0000 \\
         &   10. &   0.1650  &  0.3790 &   0.0370  &  0.0000  &  0.0000 \\
10$^{-4}$&  100. &   0.0449  &  0.0896 &   0.0128  &  0.0361  &  0.0343 \\
         & 1000. &   0.0383  &  0.0771 &   0.0107  &  0.0352  &  0.0373 \\
         &10000. &   0.0380  &  0.0764 &   0.0106  &  0.0345  &  0.0377 \\
\hline

         &    1. &   0.0026 &   0.0028 &   0.0000  &  0.0000 & 0.0000 \\
         &   10. &   0.0064 &   0.0152 &   0.0018  &  0.0000 & 0.0000 \\
10$^{-2}$&  100. &   0.0034 &   0.0049 &   0.0025  & 0.0292  & 0.0276  \\
         & 1000. &   0.0027 &   0.0041 &   0.0021  & 0.0303  & 0.0324  \\
         &10000. &   0.0026 &   0.0041 &   0.0021  & 0.0297  & 0.0331  \\
\hline
\end{tabular}
\end{table}

\begin{table}
 \caption{Similar to Table \ref{h2oyields_table} for {\htwo}. 
 For information on the channels, see Table \ref{h2channels_table}. 
 EE1 refers to EE1B+EE1C+EE1EF+EE1SS, 
 EE3 to EE3b+EE3c+EE3a+EE3e+EE3ss, and 
 IO to IO1+IO2. 
 For comparison, the normalized yield for ionization of neutral {\htwo} calculated by \citet{dalgarnoetal1999} 
\textcolor{black}{in the high-energy limit (say $E_0${$>$}1,000 eV)}
  is 1/37.6=0.0266 eV$^{-1}$ and the yield measured by \citet{combecher1980} at $E_0$=500 eV is 1/38.64=0.0259 eV$^{-1}$ \textcolor{black}{(we obtain 0.0284 eV$^{-1}$ at 500 eV)}.
  }
 \label{h2yields_table}
 \begin{tabular}{llccccc}

\hline
               &            & \multicolumn{5}{c}{Normalized yield [eV$^{-1}$]} \\
               \cline{3-7}
  $x_{\rm{e}}$ & $E_0$ [eV] & VE & EE1 & EE3 & DN & IO \\  
\hline  
   &    1. & 0.5270  &  0.0000  &  0.0000  &  0.0000  & 0.0000  \\
   &   10. & 1.0700  &  0.0000  &  0.0000  &  0.0000  & 0.0000  \\
0. &  100. & 0.1720  &  0.0228  &  0.0127  &  0.0024  & 0.0253  \\
   & 1000. & 0.1440  &  0.0241  &  0.0085  &  0.0023  & 0.0288  \\
   &10000. & 0.1390  &  0.0246  &  0.0081  &  0.0024  & 0.0288  \\
\hline

         &    1. &  0.0102  &  0.0000 &   0.0000 &   0.0000 &   0.0000  \\
         &   10. &  0.3260  &  0.0000 &   0.0000 &   0.0000 &   0.0000  \\
10$^{-4}$&  100. &  0.0484  &  0.0226 &   0.0118 &   0.0023 &   0.0251  \\
         & 1000. &  0.0395  &  0.0241 &   0.0079 &   0.0023 &   0.0287  \\
         &10000. &  0.0383  &  0.0245 &   0.0075 &   0.0024 &   0.0288  \\
\hline

         &    1. &  0.0001  &  0.0000  &  0.0000  &  0.0000  &  0.0000  \\
         &   10. &  0.0057  &  0.0000  &  0.0000  &  0.0000  &  0.0000  \\
10$^{-2}$&  100. &  0.0015  &  0.0181  &  0.0049  &  0.0019  &  0.0211  \\
         & 1000. &  0.0011  &  0.0212  &  0.0027  &  0.0020  &  0.0262  \\
         &10000. &  0.0010  &  0.0219  &  0.0026  &  0.0022  &  0.0265  \\
\hline
\end{tabular}
\end{table}

\begin{table}
 \caption{Similar to Table \ref{h2oyields_table} for {\otwo}. 
 For information on the channels, see Table \ref{o2channels_table}. 
 VE refers to VE1+VE2+VE3, EE to EEa+EEb+EEcA'A, and 
 IO to IO1+IO2+IO3. For comparison, \citet{combecher1980}
has measured normalized yields for ionization in neutral {\otwo} of 
1/44.76=0.02234 eV$^{-1}$ for $E_0$=100 eV and 
1/32.55=0.03072 eV$^{-1}$ for $E_0$=500 eV
\textcolor{black}{we obtain 0.0394 eV$^{-1}$ at 500 eV}.}
 \label{o2yields_table}
 \begin{tabular}{llcccc}

\hline
               &            & \multicolumn{4}{c}{Normalized yield [eV$^{-1}$]} \\
               \cline{3-6}
  $x_{\rm{e}}$ & $E_0$ [eV] & VE & EE & DN & IO \\  
\hline  
   &    1. & 0.0000  &  0.0000  &  0.0000  &  0.0000   \\
   &   10. & 0.4828  &  0.1429  &  0.0367  &  0.0000   \\
0. &  100. & 0.1071  &  0.0417  &  0.0335  &  0.0360   \\
   & 1000. & 0.0948  &  0.0369  &  0.0286  &  0.0399   \\
   &10000. & 0.0937  &  0.0366  &  0.0282  &  0.0403   \\  
\hline

         &    1. &  0.0000  &  0.0000  &  0.0000  &  0.0000  \\
         &   10. &  0.2731  &  0.0746  &  0.0261  &  0.0000  \\
10$^{-4}$&  100. &  0.0493  &  0.0187  &  0.0321  &  0.0358  \\
         & 1000. &  0.0446  &  0.0168  &  0.0272  &  0.0397  \\
         &10000. &  0.0437  &  0.0165  &  0.0269  &  0.0402  \\
\hline

         &    1. &  0.0000  &  0.0000  &  0.0000  &  0.0000  \\
         &   10. &  0.0087  &  0.0023  &  0.0010  &  0.0000  \\
10$^{-2}$&  100. &  0.0027  &  0.0014  &  0.0139  &  0.0292   \\
         & 1000. &  0.0024  &  0.0013  &  0.0113  &  0.0344   \\
         &10000. &  0.0024  &  0.0013  &  0.0111  &  0.0351   \\
\hline
\end{tabular}
\end{table}

\begin{figure*}
 \includegraphics[width=15cm]{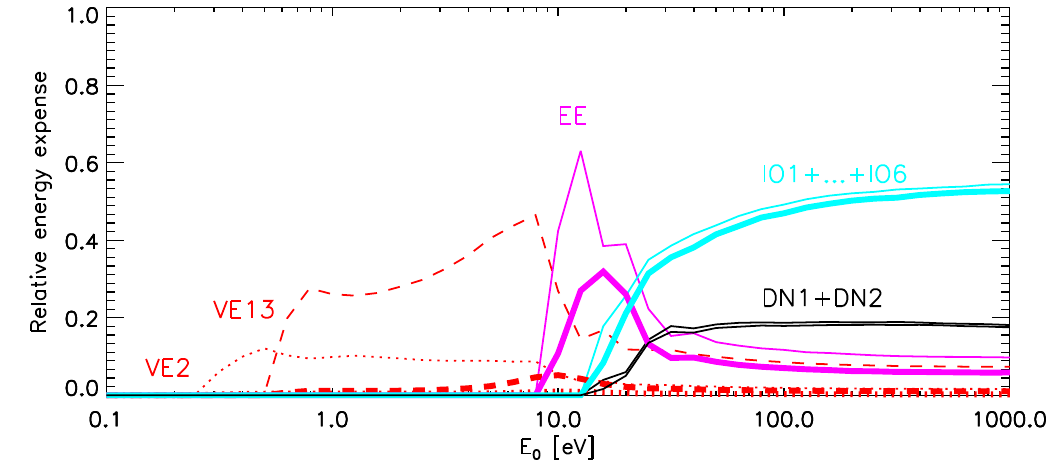}\\
 \includegraphics[width=15cm]{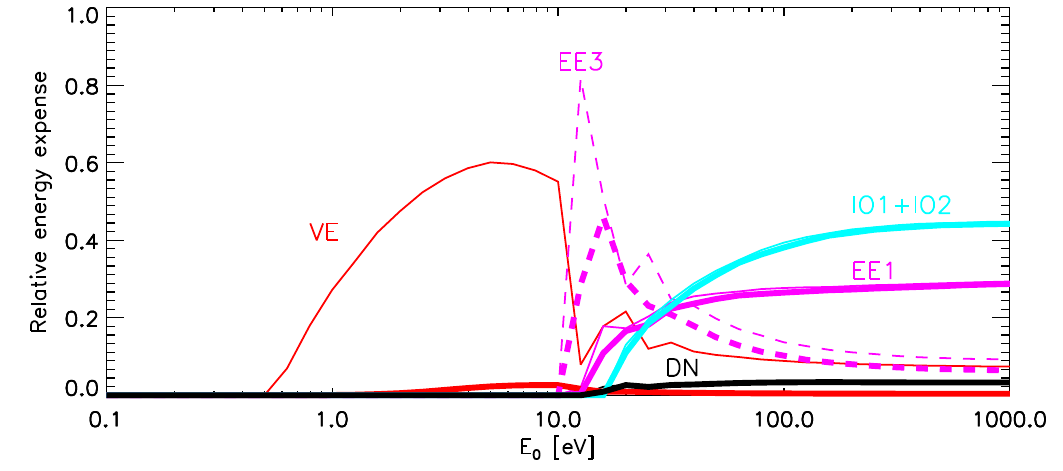}\\
 \includegraphics[width=15cm]{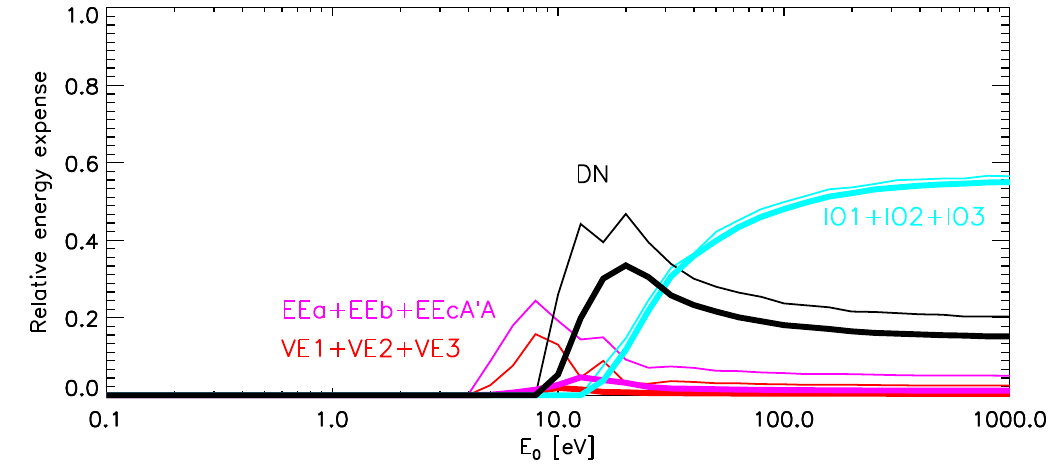}
 \caption{Some outcomes from the MC scheme. For numerical experiments in which electrons of energy $E_0$ are injected into a gas composed of a single molecule and thermal electrons, 
 fraction of that energy that is expended by the non-thermal electrons in excitation, dissociation and ionization. The top, middle and bottom panels refer to {\htwoo}, 
{\htwo} and {\otwo}, respectively.
  Thin lines are for a fractional ionization $x_{\rm{e}}$=0, and thick lines for $x_{\rm{e}}$=10$^{-3}$.
 For all three molecules, the energy expended in channels with low thresholds (typically, vibrational and electronic excitations) is very sensitive to the fractional ionization of the gas. High-energy electrons expend most of their energy in ionization, dissociation and electronic excitation; the specific partitioning is however very molecule-dependent.
  }
 \label{energyexpense_fig}
\end{figure*}

Figure \ref{energyexpense_fig} is also based on the 
MC calculations. It shows, for $x_{\rm{e}}$=0 and 10$^{-3}$, 
how a primary electron expends its initial energy $E_0$
from injection to thermalization.
The main features in this figure are consistent with the information in  the stopping cross sections. 
Essentially, ionization, dissociation into neutrals, electronic, vibrational and rotational excitation collisions progressively dominate the extraction of energy from the non-thermal electrons as their initial energy $E_0$ decreases from keV to less than 1 eV.
The fraction of 
$E_0$ that goes into channels with low energy thresholds (vibrational but also electronic excitation) is very sensitive to the fractional ionization of the gas, the reason being that electron-electron collisions become efficient at low energies.

\section{Application to the exoplanet Trappist-1 b}

\textcolor{black}{
We next demonstrate the above ideas with the atmospheric profiles of exoplanet Trappist-1 b calculated in  ref.\cite{garciamunoz2023_aa} 
(Fig. 4 in that work shows
the densities of {\htwoo}, {\otwo}, H, O, some ions and thermal electrons) 
 \textcolor{black}{ with a model based on earlier formulations of the hydrodynamic-phochemical problem in exoplanet atmospheres\cite{garciamunoz2007b,garciamunozetal2020,garciamunozetal2021}}.}
The hydrodynamic model solves the equations of the gas in a 
one-dimensional (vertical direction only) geometry, as it is heated by stellar irradiation and accelerates into space. It works on the basis  that the deep atmosphere is pure {\htwoo} 
that transitions 
into neutral atoms and then into ions by the simultaneous effects of photodissociation/-ionization, chemical kinetics \textcolor{black}{(about 150 chemical processes in total)} and the dynamics of the escaping gas.
\textcolor{black}{We take from the hydrodynamic model the rate of photoionization for each atom and molecule in the atmospheric gas and the energy of the primary electrons, as well as the temperature and number density profiles for the thermal electrons, atoms and molecules in the atmosphere. With this information, we solve the slowing down problem locally, letting the non-thermal electrons interact with the thermal electrons, the H, O and {\op} atoms, and the {\htwoo}, {\htwo} and {\otwo} molecules in the atmosphere. On output, the MC scheme provides the productions yields at each altitude. We elaborate further on these ideas in what follows.}
\\

The hydrodynamic model calculates the photoionization rate 
[$e^-$cm$^{-3}$s$^{-1}$] of $X$ as 
$J_{\nu}[X]=
\sum_i W_{0,i}=
  \sum_i [X] \sigma^{\rm{pi}} (\lambda_i) \mathcal{F}_{\lambda}(\lambda_i) \frac{\lambda_i}{hc} \Delta \lambda_i,$
where [$X$] stands for number density [cm$^{-3}$], 
$\sigma^{\rm{pi}} (\lambda_i)$ for photoionization cross section
[cm$^2$], $\mathcal{F}_{\lambda}(\lambda_i)$ for radiation flux [erg cm$^{-2}$s$^{-1}$],  and $\Delta\lambda_i$ is the spectral bin width near $\lambda_i$. 
$W_{0,i}$ is the rate at which primary electrons of energy {$E_{0,i}$}{$\sim$}$hc$/$\lambda_i${$-$}IP($X$) are released. 
Considering all atoms and molecules $X$ that photoionize, the set \{$E_{0,i}$; $W_{0,i}$\} defines the energy spectrum of the primary electrons in the atmosphere. 
\\

\begin{figure*}
 \includegraphics[width=9cm]{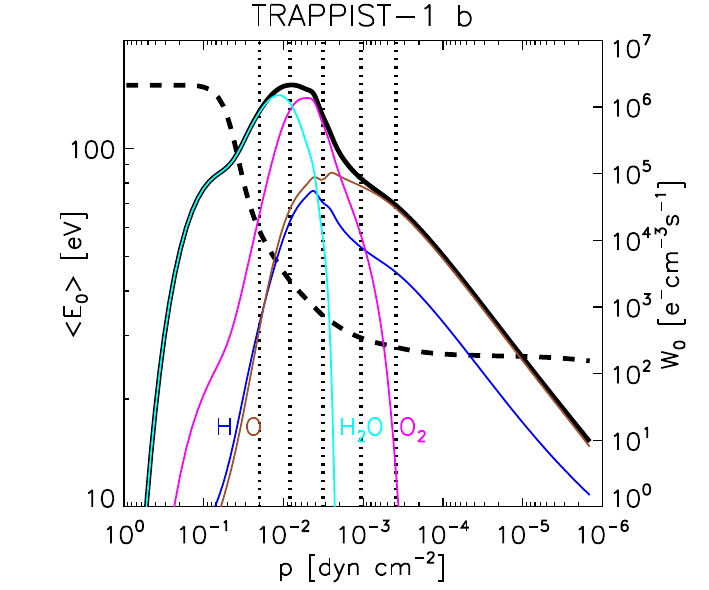}\\
 \vspace{-0.5cm}
 \includegraphics[width=9cm]{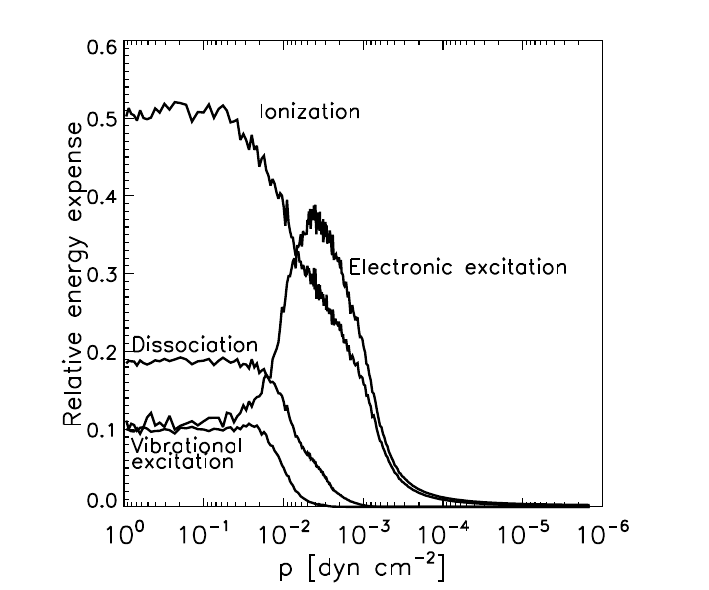} 
 \caption{Top panel. Characteristic energy {$<$}{$E_0$}{$>$} (left axis, dashed line) and production rate of primary electrons (right axis, solid lines) in the Trappist-1 b atmosphere\cite{garciamunoz2023_aa}. In color, the 
 dominating contributions to $W_0$ separated by atom and molecule.
 From left to right, the dotted lines indicate where the fractional ionization $x_e$ (number density of thermal electrons / total number density of neutrals)=10$^{-5}$, 10$^{-4}$, 10$^{-3}$, 10$^{-2}$ and 10$^{-1}$.
 Bottom panel. 
In the partial problem for the energy balance considered here,  
 fraction of the primary electrons' energy
  ($W^{E}_{\rm{chn}}$/{$<$}{$E_0$}{$>$}{$W_0$}, see text)
  that goes into ionization, dissociation and excitation. 
The energy expended in 
  rotational excitation and elastic momentum-transfer collisions is not represented; both types of collisions will result in heating at the densities of interest here.
 The fluctuations arise from statistical noise in the MC calculations. 
 The subsequent chemical kinetics and radiation problems will dictate how much of that energy is actually channeled into heating, a question that must be determined in the general problem.
  }
 \label{E0W0_trappist1b_fig}
\end{figure*}

Figure \ref{E0W0_trappist1b_fig} (top)
shows the total rate $W_0$=$\sum_i${$W_{0,i}$}
of primary electrons, plus some information on the photoionizing molecule from which the electrons are ejected,  
and the primary electrons' characteristic energy
{$<$}{$E_0$}{$>$}=$\sum_i${$E_{0,i}$}{$W_{0,i}$}/{$W_{0}$}. The variation of these quantities with altitude is governed by the gas composition, and by the spectral dependence of the photoionization cross sections and the stellar flux.
For Trappist-1 b, 
{$<$}{$E_0$}{$>$}{$\sim$}25-30 eV
in the upper layers of the atmosphere, 
the reason being that the adopted stellar spectrum\cite{wilsonetal2021} is particularly strong at wavelengths {$\sim$}300 {\AA}. 
{$<$}{$E_0$}{$>$} increases rapidly towards the deeper layers.
The limiting value of $\sim$150 eV seen in the figure
is however an artifact caused by the modest spectral resolution used  to describe the stellar spectrum.\cite{garciamunoz2023_aa} In reality, {$<$}{$E_0$}{$>$} should rise to hundreds of eV towards the pressure level of 1 dyn cm$^{-2}$.
Notwithstanding this caveat, the {$<$}{$E_0$}{$>$} profile of Fig. \ref{E0W0_trappist1b_fig} (top) suffices to demonstrate a few key points.
The figure also shows that 
at pressures $>$0.002 dyn cm$^{-2}$
the {\htwoo} and {\otwo} molecules dominate the production of primary electrons, whereas the H and O atoms dominate at lower pressures. 
This pressure level sets approximately the borderline between the molecular and atomic atmosphere.
The transition from moderate to large {$<$}{$E_0$}{$>$}
is smooth and spans a few scale heights.
\\

Used together, Figs. 
\ref{energyexpense_fig} and \ref{E0W0_trappist1b_fig} (top) help 
anticipate where in the atmosphere each 
inelastic channel becomes relevant. 
Indeed, 
because {$<$}{$E_0$}{$>$} remains always
$>$10-20 eV (the approximate value that separates the inelastic channels with low energy thresholds, namely rotational and vibrational excitations, and those with high thresholds, namely electronic excitation, dissociation and ionization;  
Fig. \ref{energyexpense_fig}), 
it is expected that collisions involving electronic excitation, dissociation and ionization will generally dominate the slowing down of the non-thermal electrons
wherever the gas remains predominantly neutral. In turn, elastic electron-electron collisions will become increasingly relevant at moderate-to-large fractional ionizations. 
These tentative predictions are refined below.
\\

Given \{$E_{0,i}$; $W_{0,i}$\}, 
 the rate [erg cm$^{-3}$s$^{-1}$] at which the primary electrons expend their energy in a given type of collision 
 is $W^E_{\rm{chn}}=
  \sum_i W_{0,i}(E_{0,i}) \Phi_{\rm{chn}}(E_{0,i})
  \Delta E_{\rm{chn}}$. 
Here, script $_{\rm{chn}}$ refers to the collisional channel,  $\Phi_{\rm{chn}}(E_{0,i})$ to the production yield, and 
$\Delta E_{\rm{chn}}$ to the energy extracted from the non-thermal electron upon collision (column before the last in Tables \ref{h2ochannels_table}-\ref{o2channels_table} for {\htwoo}, {\htwo} and {\otwo}). 
The dimensionless 
$W^E_{\rm{chn}}$/$<${$E_0$}$>${$W_0$}  conveys the fraction of the primary electrons' energy 
that goes into the channel. 
\\

Figure \ref{E0W0_trappist1b_fig} (bottom)
shows $W^E_{\rm{chn}}$/$<${$E_0$}$>${$W_0$} for 
excitation (vibrational and electronic presented separately),  dissociation and ionization. The normalization factor $<${$E_0$}$>${$W_0$} can be worked out from the top panel. 
Three regions can be distinguished, which can be rationalized by the local values of {$<$}{$E_0$}{$>$}, $x_{\rm{e}}$ and by whether the gas is in atomic or molecular form.\\

In the first region, that spans pressures $>$0.02 dyn cm$^{-2}$ and is characterized by large {$<$}{$E_0$}{$>$} values and water remains dominant, the non-thermal electrons lose most of their energy through ionization and dissociation collisions, with both electronic and vibrational excitation also contributing to some extent. The fraction of energy that goes directly into heat (including here rotational excitation, as the molecules remain in rotational LTE to very low pressures\citep{garciamunozetal2024}, and elastic collisions) is only {$\sim$}10{\%}.
In the second region, between 10$^{-2}$ and a few times 10$^{-4}$ dyn cm$^{-2}$,  
the non-thermal electrons lose most of their energy through electronic excitation 
collisions, with a significant yet somewhat smaller fraction of energy going into ionization. This region brackets the conversion of {\htwoo} into H and O atoms, and indeed all three gases contribute significantly. 
The region is characterized by modest {$<$}{$E_0$}{$>$} values, yet the fraction of energy going into vibrational excitation is very small because $x_{\rm{e}}$ is large enough that the low-energy non-thermal electrons are efficiently removed in collisions with the thermal electrons.
In the third region, that spans towards lower pressures, 
{$<$}{$E_0$}{$>$}{$\sim$}25 eV and $x_{\rm{e}}${$>$}0.1, the non-thermal electrons lose essentially all of their energy through electron-electron collisions that heat the gas.
\\

Two additional comments are due. 
Firstly, the molecular-atomic transition that occurs between the first and second regions introduced above is particularly important because that is where the 
energy injected by the primary electrons (as quantified by {$<$}{$E_0$}{$>$}{$W_0$}) peaks. Our calculations show that only a small fraction of it goes initially into heating. 
Secondly, the above calculations do not tell us how much of {$<$}{$E_0$}{$>$}{$W_0$} goes ultimately into heating the gas. This consideration is related to the distinction between the partial and general problems introduced earlier.
For example, the newly formed ions may recombine through radiative and non-radiative channels. In the first case, which is typical of atomic ions, the transformation will not heat the gas. In the second case, which is typical of molecular ions, the transformation will result in part of the ionization energy being returned to the gas as heat. 
Similar considerations can be made 
when the collisions of the non-thermal electrons lead to 
dissociated and electronically excited products. 
For vibrationally excitated products, the amount of energy that is  radiated depends on the molecule. It will be effectively zero for the homonuclear {\htwo} and {\otwo} molecules at the densities of interest here, and an amount that will depend on the local quenching efficiency and therefore on the local density and composition\cite{garciamunozetal2024} for {\htwoo}. 
\\

Related to above, 
the production rate [cm$^{-3}$s$^{-1}$] at which a channel proceeds is
$W^{P}_{\rm{chn}}=
  \sum_i W_{0,i}(E_{0,i}) \Phi_{\rm{chn}}(E_{0,i})$. 
Each channel in the MC scheme being a first-order process, $J_{e}$=$W^{P}_{\rm{chn}}$/[$X$] 
[s$^{-1}$] 
quantifies the inverse of the characteristic time for loss of $X$ through the channel. This is equivalent\cite{garciamunoz2023_icarus} to
$J_{e} = \int \sigma_{\rm{chn}}(E) \phi_e(E) dE$, 
where $\sigma_{\rm{chn}}$ is the channel-specific
electron-collision cross section, $\phi_e$ the flux of non-thermal electrons and integration is performed over the full of electron energies. We do not use the latter formulation for $J_e$, but it is noted here to facilitate the discussion. 
\textcolor{black}{We elaborate on the equivalence between our formulation and that based on $J_e$ in the SI.}
$J_{e}$ is analogous to the photoionization/-dissociation rate coefficient $J_{\nu}$ for photon collisions. 
\\

\begin{figure*}
 \includegraphics[width=9cm]{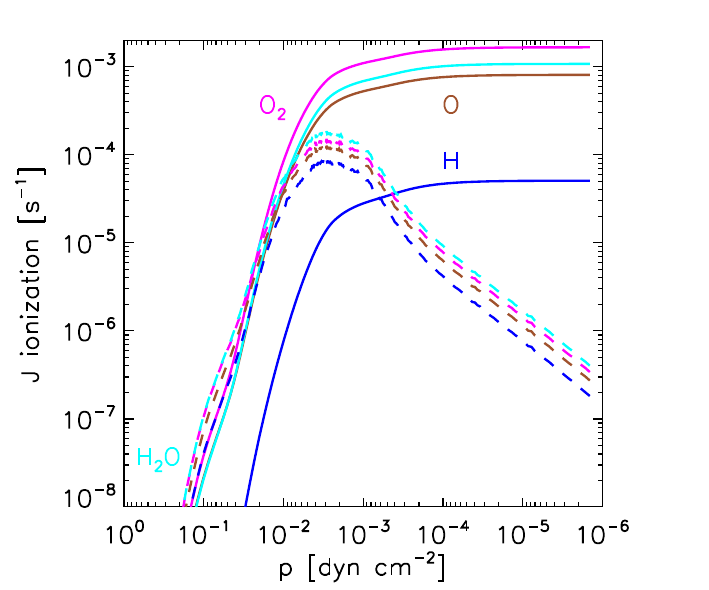}\\
 \vspace{-0.5cm}
 \includegraphics[width=9cm]{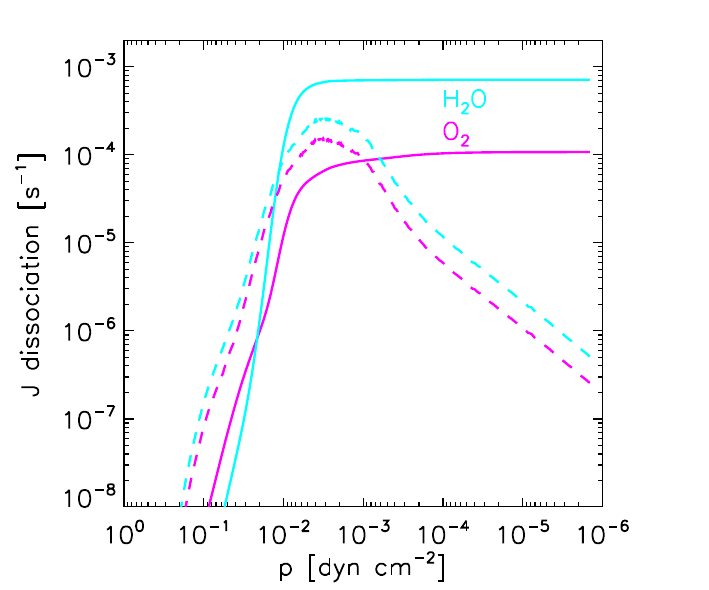} 
 \caption{Top panel. Coefficient 
 for photoionization $J_{\nu}$ (solid) and ionization by secondary electrons $J_e$ (dashed) for 
 {\htwoo}, {\otwo}, H and O in the atmosphere of Trappist-1 b.
 Bottom panel. For {\htwoo} and {\otwo}, the corresponding dissociation coefficients.
  }
 \label{Je_fig}
\end{figure*}

The top panel of Fig. \ref{Je_fig} shows the $J_e$ coefficients for ionization (all partial channels merged) 
of {\htwoo}, {\otwo}, H and O, and compares them with the photoionization coefficients $J_{\nu}$. 
The bottom panel shows $J_e$ for dissociation into neutrals of the {\htwoo} and {\otwo} molecules and compares them to the corresponding $J_{\nu}$ for dissociation. 
For {\htwoo}, we have assumed that in addition to channels DN1 and DN2, channel EE leads to dissociation with 100{\%} efficiency. For {\htwo} (negligible here and not shown), we assumed that electronic excitation into the singlet and triplet states leads to dissociation with 10{\%} and 100{\%} efficiencies, respectively (see the SI for a justification). 
For {\otwo}, only channel DN leads to dissociation. 
The last columns of Tables \ref{h2oyields_table}-\ref{o2yields_table} summarize some of the information regarding how the channels in the MC scheme contribute to ionization and dissociation.
\\

The main conclusions to draw from Fig. \ref{Je_fig} are:
\begin{itemize}
    \item For ionization, $J_e$ \textcolor{black}{can be} up to a few times larger than the $J_{\nu}$ that contributes the majority of primary electrons to the atmosphere (see Fig. \ref{E0W0_trappist1b_fig}, top panel).
\textcolor{black}{
This is particularly true at the higher pressures of the atmosphere, 
where most of the primary electrons are created from 
{\htwoo} and {\otwo}, yet the production of secondary electrons is significantly larger. 
}    
The reason is that the release of primary electrons, regardless of the identity of the photoionizing atom  or molecule, controls the 
flux of non-thermal electrons that appears in the expression for $J_e$. 
The ratio $J_e$/$J_{\nu}$ correlates with $<${$E_0$}{$>$} because as the primary electrons become more energetic, they can participate in additional ionization collisions. 
This ratio is of up to a few for {O}, {\htwoo} and {\otwo}, but much larger for H because its photoionization cross \textcolor{black}{section} drops rapidly towards the short wavelengths.

\item For ionization, the $J_e$ coefficients are comparable for atoms and molecules. Indeed, exposed to a given flux of non-thermal electrons $\phi_e$, $J_e$ is controlled by the electron-collision cross sections, which are relatively similar for the atoms and molecules being considered and is not very sensitive to the actual energy threshold.

\item For {\htwoo} and {\otwo}, the $J_e$ for dissociation into neutrals is comparable to the $J_e$ for ionization. The reason is again that exposed to a given flux of non-thermal electrons, the $J_e$ value is controlled by the electron-collision cross sections of the specific channels, and these are comparable for the molecules being considered.

\item For {\htwoo} and {\otwo}, the $J_e$ for dissociation into neutrals can be notably larger than the corresponding $J_{\nu}$. This finding may however depend on 
the FUV spectrum of the host star, which dictates the photodissociation rate coefficient in optically thin conditions. 
Indeed, one could expect that sufficiently deep in the atmosphere 
$J_e${$>>$}$J_{\nu}$ when the FUV spectrum is weak, but 
$J_e${$<<$}$J_{\nu}$ when the FUV spectrum is moderate-to-strong  \textcolor{black}{provided that it } remains unattenuated.
   
\end{itemize}

\section{Summary}

The current work demonstrates a new model for simulating the energy transfer from non-thermal electrons to the {\htwoo}, {\htwo} and {\otwo} molecules. 
It relies on solving the slowing down problem of the electrons by means of a MC scheme. 
We make available the model and tabulations of production yields for numerical experiments in which monoenergetic electrons are released into a single-molecule gas. These  will hopefully aid future investigations in astrophysics and other applied sciences. \\

We use the model to explore the fate of the primary and secondary electrons produced from photoionization in the exoplanet Trappist-1 b. 
Some of the findings should apply to more general atmospheres too.
It is found that the elastic and inelastic channels that dominate the energy transfer vary with altitude.
We rationalize those variations on the basis of the primary electrons' characteristic energy $E_0$
(which varies from $\sim$25 eV
in the uppermost atmospheric layers to $\sim$150 eV in the deeper layers), the fractional ionization of the gas $x_e$ and whether this is in molecular or atomic form.
This transition will cause variations in the heating and chemical kinetics, which we characterize partially. 
In particular, we note that 70-90{\%} of the primary electrons' energy is expended in ionizing, dissociating and exciting electronically the gas in the region where the energy deposition peaks. 
Amongst the collisional channels that are available to molecules but not to atoms, vibrational excitation may be easily attenuated by 
electron-electron collisions unless the gas remains virtually neutral;  dissociation into neutrals affects the slowing down of the high-energy primary electrons, but whether it remains competitive against FUV-driven photodissociation at breaking the molecular bonds may sensitively depend on the stellar FUV spectrum.
\\

The problem considered here reveals complex connections between the non-thermal electrons and the ensuing heating, excitation, dissociation and ionization of the atmospheric gas, which confirm the importance of 
preserving this complexity in atmospheric models. 
\textcolor{black}{
We have focused on the partial problem of calculating 
 the rates for excitation, dissociation and ionization caused by the non-thermal electrons but without including those effects in the hydrodynamical modelling.
In future work, we will investigate the general problem, thus exploring the feedbacks that occur when these effects are integrated in the full solution of the mass, momentum and energy equation of exoplanet atmospheres.
}

\clearpage
\newpage

\begin{suppinfo}

On the dissociation of {\htwo}; On the formulation of the rate coefficient $J_e$.
Link to the Monte Carlo model
and the tabulated production yields discussed in the paper for a gas composed of a molecule 
({\htwoo}, {\htwo} or {\otwo}) plus thermal electrons.
\\

\end{suppinfo}

\newpage

\begin{figure*}
 \includegraphics[width=15cm]{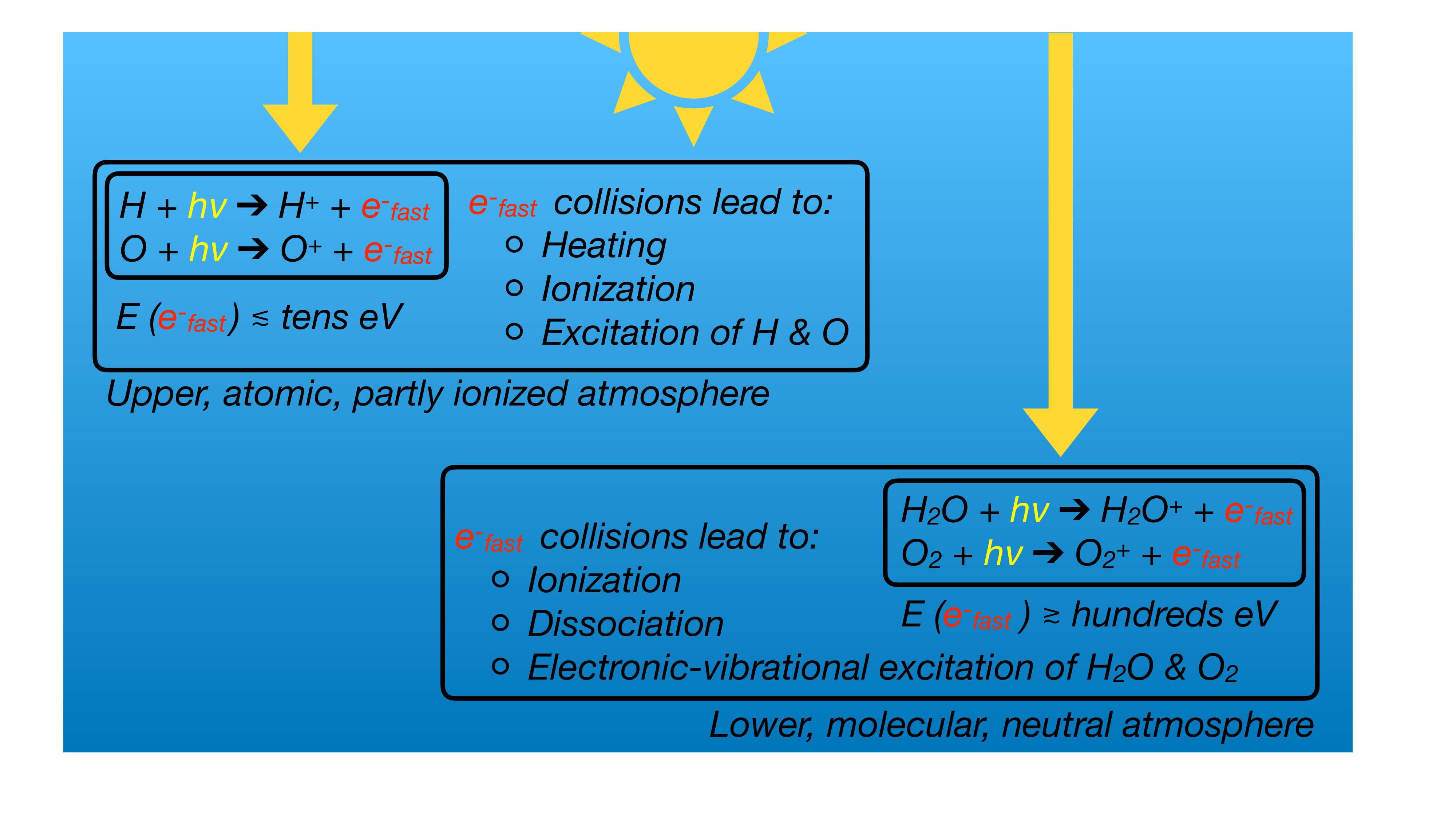} \\
 \caption{For Table of Contents Only.}
\end{figure*}

\newpage
\bibliography{achemso-demo}

\end{document}